\documentclass[12pt,preprint]{aastex}
\usepackage{emulateapj5,apjfonts}
\input psfig.tex

\newcommand{\kms}{$\rm {km}~\rm s^{-1}$}
\newcommand{\Msun}{M_\odot}

\begin{document}

\title{The Black Hole Mass, Stellar M/L, and Dark Halo in M87}
\author{Karl Gebhardt\altaffilmark{1} \& Jens Thomas\altaffilmark{2,3}}

\altaffiltext{1}{Department of Astronomy, University of Texas at
Austin, 1 University Station C1400, Austin, TX 78712;
gebhardt@astro.as.utexas.edu}

\altaffiltext{2}{Universit\"atssternwarte M\"unchen, Scheinerstra\ss e 1, 
D-81679 M\"unchen, Germany}

\altaffiltext{3}{Max-Planck-Institut fuer Extraterrestrische Physik,
Giessenbachstrasse, D-85748 Garching, Germany; jthomas@mpe.mpg.de}

\begin{abstract}

We model the dynamical structure of M87 (NGC4486) using high spatial
resolution long-slit observations of stellar light in the central
regions, two-dimensional stellar light kinematics out to half of the
effective radius, and globular cluster velocities out to 8 effective
radii. We simultaneously fit for four parameters, black hole mass,
dark halo core radius, dark halo circular velocity, and stellar
mass-to-light ratio. We find a black hole mass of
$6.4(\pm0.5)\times10^9~\Msun$ (the uncertainty is 68\% confidence
marginalized over the other parameters). The stellar
M/L$_V=6.3\pm0.8$. The best-fitted dark halo core radius is
$14\pm2$~kpc, assuming a cored logarithmic potential. The best-fitted
dark halo circular velocity is $715\pm15$~\kms. Our black hole mass is
over a factor of 2 larger than previous stellar dynamical measures,
and our derived stellar M/L ratio is 2 times lower than previous
dynamical measures. When we do not include a dark halo, we measure a
black hole mass and stellar M/L ratio that is consistent with previous
measures, implying that the major difference is in the model
assumptions. The stellar M/L ratio from our models is very similar to
that derived from stellar population models of M87. The reason for the
difference in the black hole mass is because we allow the M/L ratio to
change with radius. The dark halo is degenerate with the stellar M/L
ratio, which is subsequently degenerate with the black hole mass. We
argue that dynamical models of galaxies that do not include the
contribution from a dark halo may produce a biased result for the
black hole mass. This bias is especially large for a galaxy with a
shallow light profile such as M87, and may not be as severe in
galaxies with steeper light profiles unless they have a large stellar
population change with radius.

\end{abstract}


\section{Introduction}

M87 has been the poster child for black hole studies due to its
central jet, large central velocity dispersion, and having the largest
measured black hole mass. Sargent et al. (1978) and Young et
al. (1978) estimate a black hole mass of $5\times10^9\Msun$ using the
surface brightness and velocity dispersion profiles. Subsequently, a
large number of studies based on the stellar dynamics provide
estimates from $1-3\times10^9~\Msun$ (see review of Kormendy \&
Richstone 1995). The latest estimate from Harms et al. (1994) use {\it
Hubble Space Telescope} (HST) to provide a gas kinematic map and black
hole estimate based on gas dynamics. Macchetto et al. (1997) analyse
additional HST gas kinematics to derive a black hole mass of
$3.2(\pm0.9)\times10^9~\Msun$, which is considered the standard
estimate for M87.

As one of the nearest galaxies and one of the largest galaxies, it is
clear why M87 has had a long and important history of having it's
black hole mass measured. In fact, M87 is often used as the anchor for
the upper end of the black hole mass distribution. There are
significant consequences for understanding the value of the maximum
black hole mass. It strongly effects the parameterization of the black
hole mass correlations (e.g., Gultekin et al. 2009), the space density
of the most massive black holes (e.g., Lauer et al. 2007), the
comparison with mass derived from quasars which tend to be closer to
$10^{10}$ rather than $10^9~\Msun$ (e.g., Shields et al. 2003), among
many others.

A significant issue for stellar dynamical models of black hole in M87
is that the dark halo contribution has never been included in the
modeling. The latest model of the gravitational potential come from Wu
\& Tremaine (2006), McLaughlin (1999), Kronawitter et
al. (2000), Romanowsky \& Kochanek (2001), and Churazov et
al. (2008). Wu \& Tremaine use the globular cluster kinematics to
measure the enclosed mass assuming a two-integral distribution
function. McLaughlin use a isotropic model for M87 based on the
globular cluster kinematics and do not include a black hole. Both
Kronawitter et al. and Romanowsky \& Kochanek (2001) assume a
spherical distribution. Churazov et al. use the gravitational
potential as derived from X-ray gas and compare to that derived from
stellar dynamics. These studies find moderate agreement for the
gravitational potential at large radii from the two techniques. Our
goal in this paper is to understand possible degeneracies between
measuring the central black hole and the dark halo simultaneously. We
use orbit-based models which do not limit the form of the allowed
velocity anisotropies assuming axisymmetry. We focus on M87 since the
quality of the observational data for M87 has dramatically increased,
and it is worthwhile to analyze it again with the most up-to-date data
and dynamical models.

Our main result is that we find a black hole mass that is over 2 times
larger when including a dark halo compared to models that do not
contain a dark halo. This result is both generic in that it could be a
potential concern for many of the black hole studies to date, and
specific in that it applies the most up-to-date modeling to M87.
Humphrey et al. (2008) model NGC~4649, a galaxy similar in mass to
M87, using X-ray gas emission and find a black hole that is 2 times
larger than the previous stellar dynamical measure of Gebhardt et
al. (2003). Thus, there is a concern that black hole masses at the
upper end of the distribution may be biased.

We assume a distance to M87 of 17.9 Mpc.

\section{Data}

In the central regions of M87, we rely on the stellar light for both
photometry and kinematics. For the outer region, the stellar
kinematics only extend to about 40\arcsec. Therefore, at large radii
we rely on radial velocities of individual globular clusters for the
kinematics, also including the number density profile for the
clusters. Unfortunately, the stellar data ends around 40\arcsec\ and
the high signal-to-noise globular cluster data starts around
150\arcsec. Thus, there is a radial region where we presently have no
kinematic data.

\vskip 10pt \psfig{file=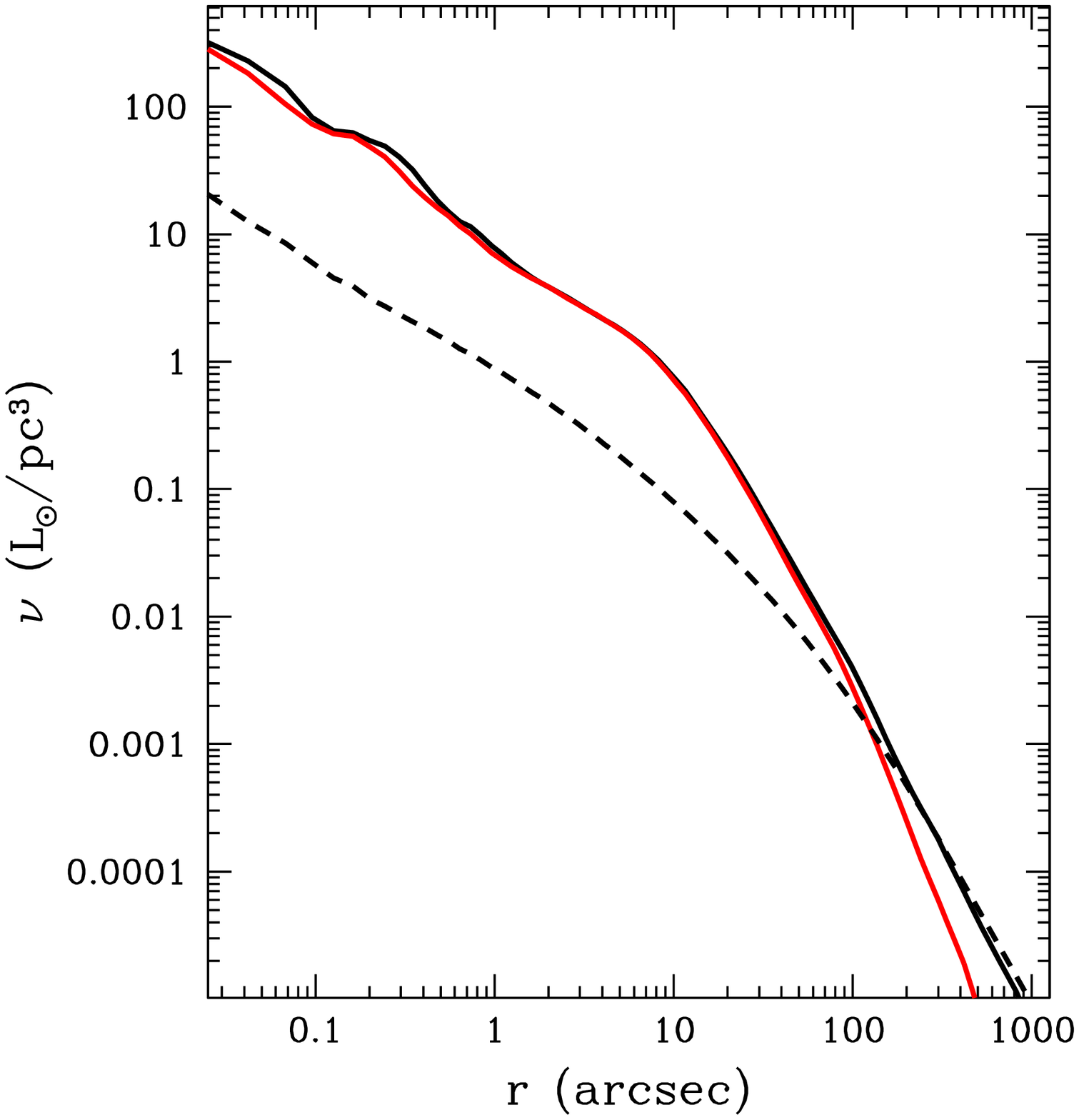,width=9cm,angle=0}
\figcaption[f1.ps]{The volume density radial profile for stars. The
black and red solid line represent the luminosity density along the
major and minor axis, respectively. These come from deprojection of
the surface brightness profile presented in Kormendy et
al. (2009). The dashed line is the deprojection of the surface number
density for globular clusters as measured by McLaughlin (1999). The
globular cluster profile is normalized, arbitrarily, to match the
stellar profile in order to compare the shapes. The actual volume
density of the clusters is much lower and their contribution to the
stellar mass is irrelevant. Furthermore, the cluster profile within
20\arcsec\ is extrapolated since there is little information to
measure a surface brightness profile. Both profiles are used in the
dynamical models for each of the components.
\label{fig1}}
\vskip 10pt

\subsection{Stellar Density}
\label{subsec:stardens}
The stellar surface density profile comes from Kormendy et al. (2009),
which is a combination of HST data from Lauer et al. (1992) and
various ground-based observations. It extends from 0.02\arcsec\ to
2400\arcsec\ (5 orders of magnitude). The surface brightness profile
in the $V$-band, ellipticity, and position angle versus radius are
shown in Kormendy et al. (2009). The ellipticity is near zero for most
of the radial range, but then increases at both small and large
radii. The increase at large radii has been known for some time (e.g.,
Weil et al. 1997), and is associated with the large rotation seen
there (Kissler-Patig \& Gebhardt 1998).  The increase at small radii
has been noted by Lauer et al. (1992) and appears to be real (as
opposed to noise). In fact, the axisymmetric deprojection of the
surface brightness to the luminosity density requires the increase at
small radii. We deprojected the surface brightness using Magorrian
(1999). This deprojection allows for a radial variation of the
observed ellipticity. Figure~1 plots the luminosity density along the
major and minor axis versus radius. The dynamical models use this
non-parametric model of the luminosity density. From this figure one
can see the increase in the ellipticity in the central region and in
the outer region (the minor axis is lower than the major axis). From
about 2--70\arcsec\ M87 is nearly round in projection. The full radial
range used in the dynamical models spans $10^7$ in the luminosity
density.

\subsection{Stellar Kinematics}
\label{subsec:starkin}

There have been many stellar kinematic studies of M87 and we rely on
the two best studies for our models. These are the kinematics from
SAURON (Emsellem et al. 2004) and from van der Marel (1994).

The SAURON Team provide two-dimensional kinematic coverage of M87 out
to about 40\arcsec. The kinematics are parameterized with the first
four coefficients of a Gauss-Hermite polynomial expansion, with
uncertainties. The two-dimensional coverage of SAURON provides the
best possible coverage for determining the stellar orbital structure
and mass-to-light profile. Our dynamical models fit the line-of-sight
velocity distribution (LOSVD) directly as opposed to fitting to
moments or Gauss-Hermite coefficients. Thus, we have to convert the
SAURON velocity moments into LOSVDs; we do this by creating Monte
Carlo simulations based on the uncertainties of the Gauss-Hermite
coefficients. This procedure assumes that the moments are uncorrelated
(further discussed in Section 4). The spatial resolution of the
dynamical models is coarser than the SAURON data in the outer regions
and similar in the inner regions (cf. Sec.~\ref{subsec:stellarmods}).
Where necessary, we rebin the SAURON data by averaging weighted by the
uncertainties over all SAURON LOSVDs in one model bin. We do not
average the Gauss-Hermite coefficients. Ideally, we would not apply
any spatial binning and use each ground-based element independently
(even though the kinematic data was already binned originally), but
this fine binning would take a significant amount of computing
resources as to make the study unfeasible. We discuss the binning
further in Sec~\ref{subsec:stellarmods}.

The kinematic data with the best spatial resolution come from van der
Marel (1994). They have 0.6\arcsec\ FWHM seeing with a 1\arcsec\
slit. The data extend to 10\arcsec. Van der Marel provides the first
four Gauss-Hermite moments and we transform to LOSVDS as we did with
the SAURON data.

Another important dataset comes from Sembach \& Tonry (1996), who
obtain kinematics in M87 out to a radius of 130\arcsec. We do not use
their kinematics mainly since they only provide velocity and velocity
dispersion, without higher order moments. Since the SAURON data are
such higher S/N and provide the first 4 moments of the LOSVD and since
our modeling code fits LOSVDs (and not moments), in order to include
the Sembach \& Tonry data we would have to mock up the higher order
moments, which would lead to additional uncertainty. Furthermore,
there appears to be an offset in the velocity dispersion from Sembach
\& Tonry compared to other kinematic studies as discussed in
Romanowsky \& Kochanek (2001). Future datasets at these radii in M87
using the full LOSVD will be very important for additional analysis.

There are many other kinematic datasets, but the S/N and areal
coverage of the SAURON data is superior to all other studies that it
will make little difference, if any, to the dynamical models. Given
there are often systematic differences between various datasets, it is
better to use just one dataset. We include the van der Marel data
since these add significantly to the recovery of the black hole
mass. The two datasets are consistent with and show no biases between
each other at the radii where they overlap.

\vskip 10pt \psfig{file=f2.ps,width=9cm,angle=0}
\figcaption[f2.ps]{The grid lines represent the spatial binning used
for the dynamical models. The points are the locations of the 278
globular cluster radial velocities. The black and red points represent
clusters on opposite sides of the minor axis, and the red points have
had their velocity flipped relative to the systemic. The colored
spatial bins are the binning used for those particular
cluster. Clusters outside those highlighted bins are not used in
the models. There are 275 cluster velocities used for the analysis.
\label{fig2}}
\vskip 10pt

\subsection{Globular Cluster Data}
\label{subsec:gckin}

The surface density profile for the globular clusters come from
McLaughlin (1999), who compile the profile based on various
datasets. We deproject this profile using a non-parametric spherical
inversion as described in Gebhardt et al. (1996). In Figure~1, we
include deprojected, normalized number density profile of the globular
clusters; we use an arbitrary scaling for the cluster profile in order
to compare the profile shapes of the clusters and stars. The total
mass of the clusters is irrelevant compared to the stars. The
dynamical models only use the slope of the luminosity density in order
to constrain the enclosed mass. At large radii the globular cluster
profile is shallower than the stellar profile, thus it is important to
use the cluster profile in the dynamical models. Within 20\arcsec\
there is very little information on the globular cluster profile and
we must use an extrapolation there. The dashed line in Figure~1 shows
what we used for the models. We have also tried 2 other
extrapolations: one assumes a flat profile inside of 10\arcsec\ and
the other has a slope inside of 20\arcsec\ that is equal to the
stellar density.  These two profiles give nearly identical results as
the adopted profile.

Globular cluster velocities are reported in Cohen (2000), Cohen \&
Ryzhov (1997) and Hanes et al. (2001). These are compiled in Cote et
al. (2001).  We use the same cut as defined in Cote et al. to remove
foreground and background contamination, based on velocity, color, and
magnitude. These cuts result in 278 velocities.

We include the individual velocities in the dynamical models by
measuring the LOSVD in different spatial regions. We divide the 278
velocities into 11 spatial bins. These spatial bins are shown in
Figure~2, where we plot the model grids. Due to the nature of the
spatial binning for the dynamical models, some bins will have too few
clusters to be useful. At radii between 40 and 140\arcsec\ the density
of clusters is low enough that we include larger bins in angle and
radius than for the outer bins. Of the 278 velocities, we then use
275.

Due to the uncertain number density profile for the globular clusters
in the central regions, we have run models including and excluding
clusters inside of 170\arcsec. There are 68 cluster velocities in this
range. Inside of 20\arcsec, there are so few clusters known (since the
galaxy light begins to dominate over the cluster light), that we
already have to extrapolate the number density profile. Inside of
100\arcsec, the number density is better known, but since the light
profile for M87 is so shallow, deprojection errors can lead to a large
effect. There are no significant changes to the besst-fitted values
between including or excluding the central cluster velocities, but the
uncertainties improve when we use more data. The largest difference,
which we discuss later, is that the dark halo scale radius is much
better measured.

From the approximately 25 velocities per spatial bin, we measure an
LOSVD and uncertainties. The LOSVDs use an adaptive kernel density
estimate adapted from Silverman (1986), and explained in detail in
Gebhardt et al. (1996). The 68\% confidence bands for the LOSVD come
from bootstrap resamplings of the data. Another option would be to
calculate moments from the individual velocities and compare the
models to the moments. Since the dynamical models fit LOSVDs directly,
we choose to use the measured LOSVDs. The moments derived from the
measured LOSVDs compare very well with the moments derived from the
individual velocities.

Since we are using an axisymmetric model, the velocities are folded
along the major axis and flipped about the minor axis (the velocity
relative to the systemic velocity switches sign when flipped to the
opposite side of the minor axis), such that all data are placed in
one quadrant. This folding preserves any rotation or angular
differences at a given radius. Rotation for the globular clusters is
significant at large radii (Kissler-Patig \& Gebhardt 1998, Cohen
2000, Cote et al. 2001) and needs to be included for a proper
dynamical model.

\begin{figure*}[b]
\centerline{\psfig{file=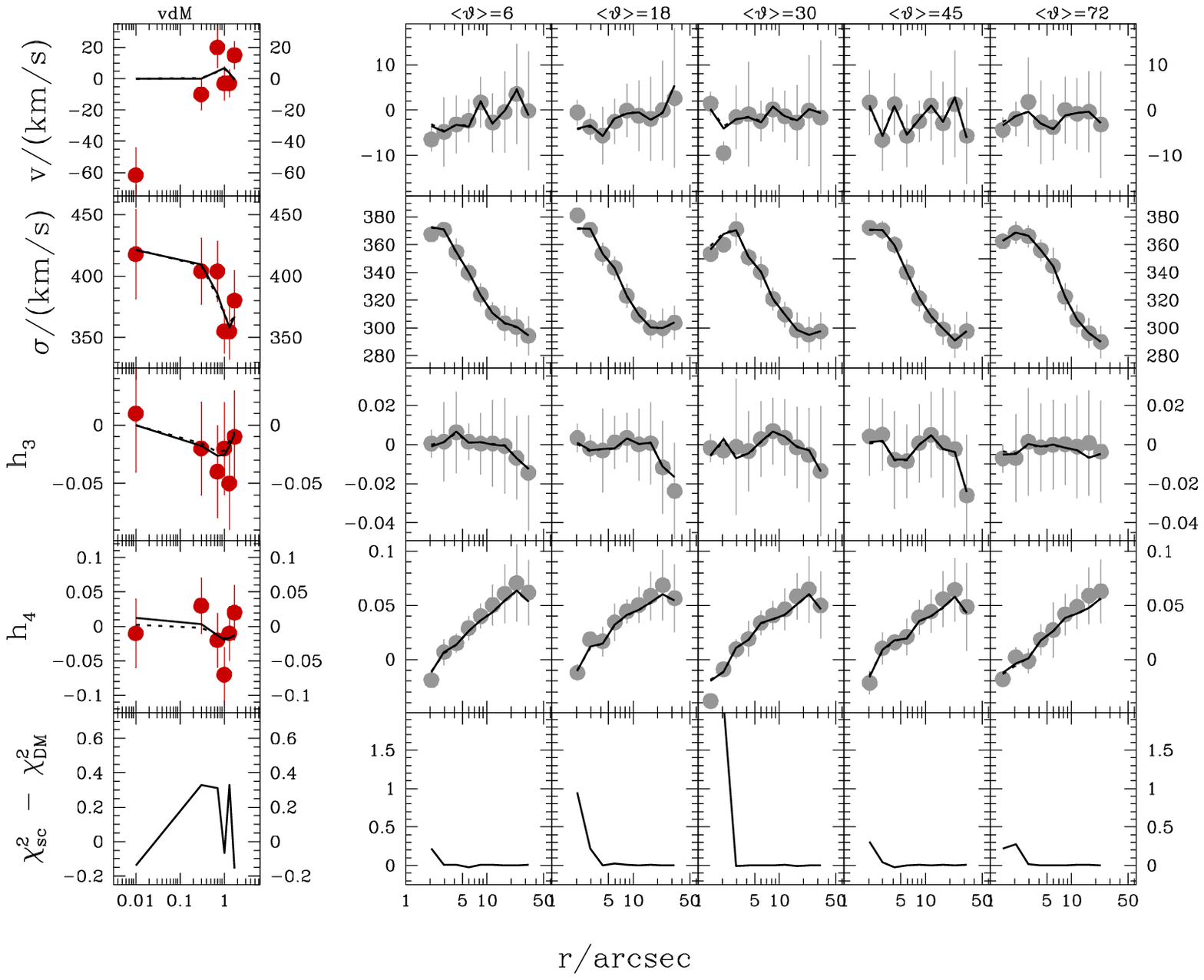,width=15cm,angle=0}}
\figcaption[data_model.eps]{Velocity $v$, velocity dispersion $\sigma$
and Gauss-Hermite coefficients $h_3$, $h_4$ as a function of
radius. Points with errors bars: observational data from van der Marel
(red; left column) and SAURON (grey; columns on the right-hand side;
each column corresponds to one angular bin of the models; the mean
lattitude is quoted on top); solid lines: best-fit model with dark
matter halo; dotted lines: best-fit model without dark matter.  Bottom
row: $\chi^2$ difference between the best-fit with dark halo
($\chi^2_\mathrm{DM}$) and without ($\chi^2_\mathrm{sc}$).  Positive
values in the bottom row indicate that the model with dark matter fits
better.
\label{data_model}}
\end{figure*}

\section{Dynamical Models}

The orbit-based models used here are described in Gebhardt et
al. (2000, 2003), Siopis et al. (2009) and Thomas et al. (2004, 2005),
and are based on the idea presented in Schwarzschild (1979). Similar
orbit-based models are presented, for example, in Rix et al. (1997),
van der Marel et al. (1998), Cretton et al. 1999, Valluri et
al. (2004). The modeling procedure consists of five steps which are
further detailed below: (i) the surface-brightness distribution is
deprojected into a 3d luminosity profile; (ii) a trial gravitational
potential is calculated including -- in our case here -- the
contribution from the stars, a dark matter halo and a black hole;
(iii) a large set of orbits is run in the trial potential; (iv) an
orbit superposition model of the galaxy in the trial potential is
constructed that matches the kinematical data as good as possible
subject to the constraints given by the 3d light profile and the trial
potential; (v) steps (i)-(iv) are repeated for a large variety of
trial mass profiles. A $\chi^2$-analysis then determines the best-fit
model and its uncertainties.

These models have undergone extensive tests for both recovery of the
dark halo and the black hole mass. Thomas et al. (2005) describe the
ability to recover the dark halo properties.  Gebhardt (2004) and
Siopis et al. (2009) describe the accuracy on the recovery of the
black hole mass and the stellar M/L ratio. The orbit-based models show
no significant bias in the recovery of the parameters for axisymmetric
systems.

\subsection{Model assumptions about the mass distribution in M87}
\label{subsec:mass87}
We calculate two different sets of dynamical models for M87 based on
different assumptions about its mass structure.

In a first set of models we do not consider a dark matter halo
explicitly.  In fact, we assume that the mass-to-light ratio of the
galaxy is spatially constant (except at the origin):
\begin{equation}
\label{eq:nohalo}
\rho = \Upsilon \, \nu + M_\mathrm{BH} \, \delta(r),
\end{equation}
with $M_\mathrm{BH}$ being the mass of the central supermassive black
hole and $\nu$ being the 3d light profile.  If it would be true that
there is no dark matter around M87 and if the mass-to-light ratio of
the stars in M87 would be constant throughout the galaxy, then
$\Upsilon$ would be our trial value for the stellar mass-to-light
ratio. However, in fact we do expect dark matter around M87 and
$\Upsilon$ from equation (\ref{eq:nohalo}) is supposed to be generally
larger than the true stellar mass-to-light ratio, because it will also
account for the potential dark mass.

In our second set of models we assume that the mass distribution
$\rho$ of M87 can be written as the sum of (1) the stellar mass
density, (2) a dark matter halo and (3) a central black hole:
\begin{equation}
\label{eq:halo}
\rho = \Upsilon \, \nu + \rho_\mathrm{DM} + M_\mathrm{BH} \,
\delta(r).
\end{equation}
The dark halo is included as described in Thomas et al. (2005). We use
both a logarithmic halo with a density profile given as:
\begin{equation}
\rho_\mathrm{DM}(r) \propto {\rm v}_c^2 \frac{3r_c^2+r^2}
{(r_c^2+r^2)^2},
\end{equation}
and a NFW potential as given by:
\begin{equation}
\label{nfw}
\rho_\mathrm{DM}(r,r_s) \propto \frac{1}{(r/r_s)(1+r/r_s)^2}
\end{equation}
(Navarro, Frenk \& White 1996).  The logarithmic halo features a flat
central density core of size $r_c$ and an asymptotically constant
circular velocity ${\rm v}_c$.  The NFW profile diverges like $r^{-1}$
towards the center and drops off with $r \propto r^{-3}$ in the outer
parts, steeper than the logarithmic halo. Concentration $c$,
scale-radius $r_s$ and the virial radius $r_v$ of NFW halos are
related via $c = r_v/r_s$.

\subsection{Modelling the stellar kinematical data}
\label{subsec:stellarmods}
For the dynamical models of the stellar kinematics we use a spatial
binning with $N_r = 28$ radial and $N_\vartheta = 5$ angular bins,
respectively (cf.  Gebhardt et al. 2000 for more details on the
spatial grids).  Figure~2 plots the radial and azimuthal plot of the
spatial binning. The model bins are generally larger than the SAURON
bins and, in that case, we average the SAURON LOSVDs as described in
Sec.~\ref{subsec:starkin}. After the spatial binning we have $N_{\cal
L}^\mathrm{stars} = 6 + 46$ LOSVDs (van der Marel \& SAURON). Each of
these stellar LOSVDs ${\cal L}^\mathrm{stars}$ is sampled by
$N_\mathrm{vel} = 19$ velocity bins.

For each trial potential (cf. equations \ref{eq:nohalo} and
\ref{eq:halo}) we determine the weighted orbit superposition that has
the lowest
\begin{equation}
\chi^2_\mathrm{stars} = \sum_{i = 1}^{N_{\cal L}^\mathrm{stars}}
\sum_{j = 1}^{N_\mathrm{vel}} \left( \frac{{\cal
L}_{ij}^\mathrm{stars}-{\cal L}_{ij}^\mathrm{mod}[\nu]}{\Delta{\cal
L}_{ij}^\mathrm{stars}} \right)^2.
\end{equation}
Here, ${\cal L}_{ij}^\mathrm{mod}[\nu]$ is the model LOSVD in spatial
bin $j$. During the $\chi^2$-minimization, the orbit model is forced
to reproduce the 3d luminosity profile $\nu$ with machine
precision. This ensures that the final model is self-consistent,
i.e. that the generated dynamical model supports those orbits out of
which it is constructed.

When two orbits differ in their spatial shape only on a sub-bin scale,
then they are equivalent with respect to the density constraints, but
their projected kinematics may be different. Then, a slight change in
the potential parameters can be compensated for by shifting light from
one orbit to the other, without affecting the density
constraints. That our model bins are larger than the SAURON data
therefore likely increases the uncertainties of the derived mass
parameters with respect to the optimum that could have been achieved
given the observational data. However, the overall mass budget of an
axisymmetric dynamical model is determined (1) by the density
constraints and (2) by the total amount of kinetic energy along the
line-of-sight (Thomas et al. 2007a). Since Fig. 3 shows that there is
no significant variation of the velocity dispersion over any of our
model bins, the total kinetic energy along the line-of-sight is
conserved after the re-binning. Furthermore, Siopis et al. (2009) and
Thomas et al. (2005) perform tests of recovery of black hole mass and
dark halo profile for analytic models. Using the same modelling code
as used here and a similar binning scheme, they find accurate recovery
of the input parameters. Thus, we do not expect that the chosen
spatial binning used here introduces a significant bias to the mass
recovery.

\begin{figure*}[b]
\centerline{\psfig{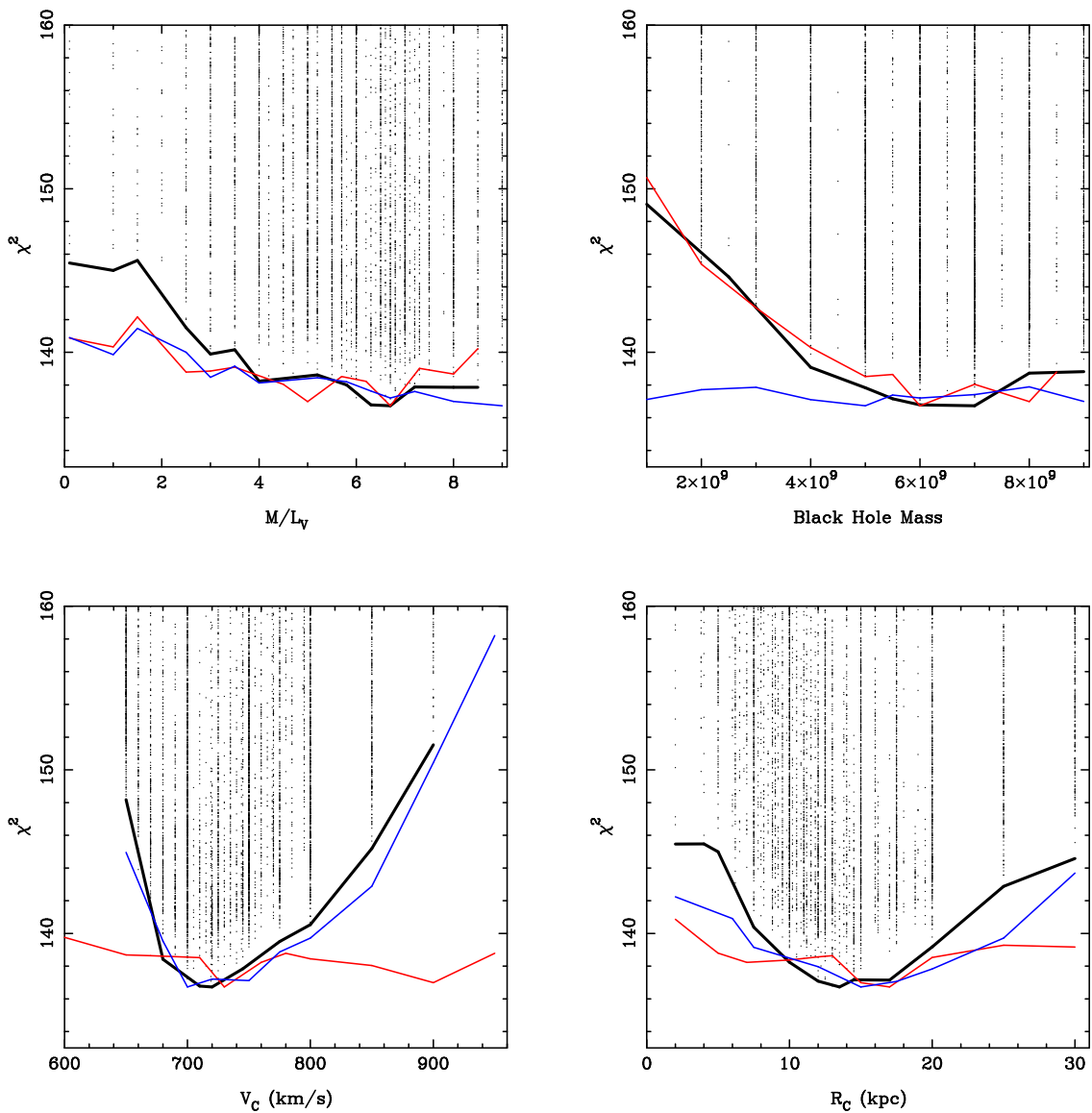}}
\figcaption[f4.ps]{$\chi^2$ versus the four parameters. In each panel,
we plot all models versus one of the parameters. A trace along the
bottom ridge thus represents the marginalized $\chi^2$ distribution,
which we use to determine the best fit and uncertainties. The small
points represent all models for the logarithmic potential. The solid
lines are the marginalized $\chi^2$ values for three models. The red
line is the $\chi^2$ determined from stars only; the blue line comes
from clusters only; the black line is from the combination. We note
that the black line is not a sum of the red and blue lines. The value
of the $\chi^2$ (the vertical axes) are for the combined value, and
that for stars and clusters have been shifted in order to see detail
in the $\chi^2$ contours. In the plot, for the stars, we add 26.3 to
the $\chi^2$, and for the clusters we add 114.2. The differential
$\chi^2$ values are preserved.
\label{fig4}}
\end{figure*}

\begin{figure*}[t]
\centerline{\psfig{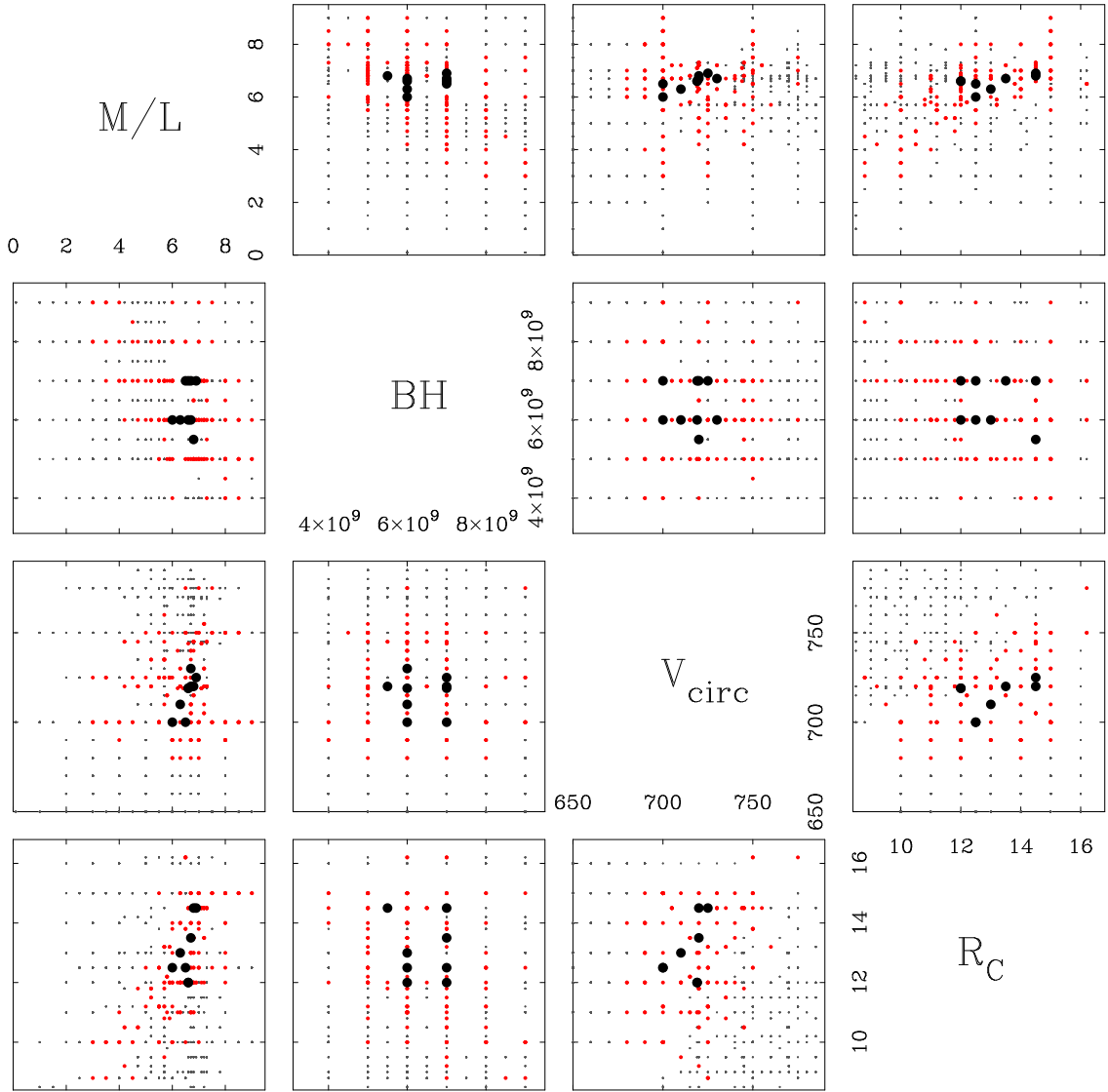}}
\figcaption[f5.ps]{Plots of the four parameters against each other,
designed to show possible degeneracies. The parameters are: M/L in the
V-band, black hole mass in $\Msun$, circular velocity of a logarithmic
halo in \kms, and core radius of the logarithmic halo in kpc. The
plotting range is chosen to highlight the region around the best fit
for each parameter. The small grey points represent the locations of
all the models; there are many additional models beyond the plotting
range that all provide significantly worse fits.  The large black
points are the locations of those models that are within the 68\%
confidence limit after marginalizing over the other parameters (i.e.,
$\Delta\chi^2=1.0$). The red points are those that are within the 95\%
confidence band. The only panel that shows an obvious correlation is
the one with core radius against M/L (the upper-right plot), which is
due to the trade between the mass of dark matter and stars. The
correlation between stellar M/L and black hole mass is weak within the
68\% bands shown here, but becomes significant for the 95\% confidence
band.
\label{fig5}}
\end{figure*}

\subsection{Modelling the globular cluster kinematics}
\label{subsec:gcmods}

The globular cluster system can have a different orbital structure
than the system of the stars, such that we need to model the globular
cluster velocities separately from the stars. However, the models for
the GC system are very similar to the models of the stellar
kinematics. In particular, we use the same mass models (cf. equation
\ref{eq:halo}, we do not attempt to model the GC kinematics without a
dark matter halo). The only difference is that the weighted orbit
superposition in each trial potential is determined by minimizing
\begin{equation}
\chi^2_\mathrm{GC} = \sum_{i = 1}^{N_{\cal L}^\mathrm{GC}} \sum_{j =
1}^{N_\mathrm{vel}} \left( \frac{{\cal L}_{ij}^\mathrm{GC}-{\cal
L}_{ij}^\mathrm{mod}[n_\mathrm{GC}]}{\Delta{\cal L}_{ij}^\mathrm{GC}}
\right)^2,
\end{equation}
where ${\cal L}^\mathrm{GC}$ are the $N_{\cal L}^\mathrm{GC} = 11$
globular cluster LOSVDs from Sec.~\ref{subsec:gckin} and
$n_\mathrm{GC}$ is the deprojected number-density profile of the
GCs. When minimizing the above $\chi^2_\mathrm{GC}$, we force the
models to reproduce $n_\mathrm{GC}$ with machine precision.

In contrast to the stars, we treat the GCs of M87 as a system of
massless test particles.  Therefore, in principle we do not need to
reproduce $n_\mathrm{GC}$ with machine precision, as any mismatch in
$n_\mathrm{GC}$ wouldn't have any effect on the potential. However, we
have tried three profiles for $n_\mathrm{GC}$ with different central
slopes and find nearly identical results. As discussed earlier, the
more important aspect is whether we include the globular cluster
velocities inside of 170\arcsec\ where the density of observed
velocities begins to decrease. The analysis presented here use all of
the central globular cluster velocites. However, models that exclude
the central velocities show no change in the best-fitted values, but
increase the uncertainties. The largest change is for the dark matter
scale radius, since when including the central velocities, we find an
increase in the uncertainty by about 50\%.

\subsection{Constraining the mass density in M87}
\label{subsec:allmods}

In total we fit for four sets of models: 1) we fit only for the black
hole mass and stellar $\Upsilon$ including the SAURON and van der
Marel data only (no globular cluster data and assuming no dark halo),
which mimics what has been done for most black hole mass studies, 2)
we fit black hole mass, stellar $\Upsilon$ and two dark halo
parameters using the SAURON data and van der Marel data only, 3) we
fit black hole mass, stellar $\Upsilon$ and two dark halo parameters
using the globular cluster data only, and 4) we fit black hole mass,
stellar $\Upsilon$ and two dark halo parameters using the SAURON data,
van der Marel data, and the globular cluster kinematics. For the last
set of models we combine the constraints from the stellar and the GC
kinematics by minimizing the sum $\chi^2_\mathrm{stars} +
\chi^2_\mathrm{GC}$. Note that because the stars and the GCs sample
different spatial scales of M87, the best-fit models 2), 3) and 4) are
not the same.

For the models 1) without dark matter halo we have to determine two
free parameters: $\Upsilon$ and $M_\mathrm{BH}$ (cf. equation
\ref{eq:nohalo}).  However, when including a dark matter halo the
amount of parameter coverage (with four parameters and two sets of
potentials) is large. We start with a uniform, sparse grid in this
four dimensional parameter space. Then we iteratively sample on a
finer grid, focussing on the region(s) around the smallest
$\chi^2$. Specifically, during iteration $n$ we determine all models
closer to the best-fit than $\Delta \chi^2 \le 1$. We then extend the
sampling around these models by filling the 80 adjacent grid points
around each of them. In this way, iteration $n$ closes the surface in
parameter space around all models with $\Delta \chi^2 \le 1$ of
iteration $n-1$. After the last iteration our fine grid covers the
entire $\Delta \chi^2 \le 1$ region. We also finely sampled a region
that is not at the minimum to assure that the sampling is adequate in
all areas of $\chi^2$; this denser sampling can be seen in the figures
described below, and we find that our sampling procedure adequately
represents the minimum $\chi^2$ contours over the four parameters.

We have run over $25,000$ orbit libraries with about $25,000$ orbits
per library, and each model takes approximately 1.5 hours of cpu. We
use the Texas Advanced Computing Center (TACC) at the University of
Texas at Austin, where they have a cluster of over 5000 nodes.

\section{Results}

We first discuss models for M87 excluding the contribution from a dark
halo.  All black holes measured to date exclude the contribution from
a dark halo (e.g., Gebhardt et al. 2003), so these models provide a
good comparison. The best-fitted M/L$_V$ is $10.2\pm0.4$, and the best
fitted black hole mass is $2.3(\pm0.6)\times10^9~\Msun$. Both of these
values are consistent with previous dynamical determinations (as
compiled in Kormendy \& Richstone 1995, Cappellari et al. 2006). The
model is compared to the data in Fig. 3 (dotted lines). It provides a
good fit to the data.

Including the dark halo significantly increases the parameter space
that has to be explored.  Given that we run over 10000 models with
four parameters it is difficult to visualize the $\chi^2$ distribution
with a single plot. Figure~4 plots the $\chi^2$ versus each of the
four parameters, including all values of the other 3 parameters. For
each panel the contour that follows the envelope of minimum values is
the marginalized $\chi^2$, and is what we use for estimation of the
parameters and their uncertainties.

\subsection{$\chi^2$ Analysis}

The combined $\chi^2=\chi^2_\mathrm{stars} + \chi^2_\mathrm{GC}$,
which we use to constrain the gravitational potential. Figure~4 plots
the $\chi^2$ values for the stars only, globular clusters only and for
the combined sample. The red line in the figure is for the stars
($\chi^2_\mathrm{stars}$); the blue line for the clusters
$\chi^2_\mathrm{GC}$); the black line for the combination. The values
on the y-axis are the sum of $\chi^2$ for the stars and clusters. In
order to plot the individual $\chi^2$ from the stars and clusters, we
must vertically shift their values; since we only use an additive
shift, the relative $\chi^2$ is preserved. A value of 26.3 is added to
the $\chi^2$ for the stars, and 114.2 for that of the clusters. We note
that the black line is not the sum of the blue and red lines, since
the model with the smallest $\chi^2$ in the sum will not be the same
model for the one that gives the minimum when using either stars or
clusters only. For the stars, with 52 LOSVDs that we use in the models
and since each are generated from 4 Gauss-Hermite parameters, the
total number of parameters is 208. For the clusters, there are 11
LOSVDs, with about 4 independent values, adding another 44
parameters. Thus, the reduced $\chi^2$ is about 0.65. This number is
similar to what we find for orbit-based models (Gebhardt et al. 2003),
which may be due to a correlation among the Gauss-Hermite coefficients
(e.g., Houghton et al. 2006).

It is clear from Figure~4 that the stars are primarily responsible for
determining the black hole mass and that the clusters are responsible
for determining the circular velocity of the dark matter. The stellar
M/L comes primarily from the stars, but it has a significant
degeneracy with the dark halo (see below). The scale radius of the
dark halo is actually determined from both stars and clusters.

Figure~4 provides the values of the best-fitted parameters and
uncertainties. For the uncertainties we use $\Delta\chi^2=1$ for each
parameter when marginalizing over the other three parameters. We
determine the range of those models within the 68\% limit and use the
middle of the range as the best-fit and half of the range as the
uncertainty. For black hole mass we find
$6.4(\pm0.5)\times10^9~\Msun$; for stellar M/L$_V$ we get
$6.3(\pm0.8)$; for logarithmic halo circular velocity we find
$715(\pm15)$~\kms; for dark halo scale radius we get
$14.0(\pm2.0)$~kpc. Our best-fit model with dark matter halo is
compared to the data in Figure~3 (solid lines). In the outer parts,
models with and without dark matter fit almost equally well to the
SAURON data, reflecting that the information about the halo comes
mostly from the GC kinematics. However, the model with dark matter
halo fits better in the inner regions (both to the van der Marel and
the inner SAURON data). This indicates that there is a significant
change in the dynamical mass-to-light ratio from $M/L_V \approx 10$ at
$r=40\arcsec$ to $M/L_V \approx 6$ at $r \approx 3\arcsec$. Towards
the very center, the two models again become very similar, which is
(1) due to the large uncertainties in the van der Marel data and (2)
due to the degeneracy between stellar $M/L$ and black-hole mass.

Our results could be biased by the choice of the dark halo
parameterisation. In particular, since we use a profile with a flat
central density core, our models are entirely dominated by luminous
mass in the inner regions. Then, in the center the luminous mass is
degenerate with the black-hole mass and, in the outer parts, the
luminous mass is degenerate with the halo (see also the discussion
below), but there is no direct communication between halo and
black-hole. The situation would be different when using a cuspy halo
profile, which could lead to models with a significant amount of dark
mass near the center, such that halo and black-hole mass could be
directly degenerate.  To check for this, we have also run models for
NFW. The grid used for the NFW models is rather coarse (we use about
1/10 of the sampling used for the logarithmic potential) and a finer
grid would be more informative, but the general shape of $\chi^2$
versus the four parameters is similar. It turns out that realistic NFW
fits (that fit the GC kinematics) are not centrally concentrated
enough to dominate the mass in the inner regions and, thus, the mass
profiles we get with NFW fits are very similar to those obtained with
logarithmic halos. (Note also, that the central logarithmic slope of
the deprojected light-profile is close to $-1$ and, thus, comparably
steep as the cusp of the NFW core). That fits with NFW and logarithmic
halos yield similar results has also been found in other early-type
galaxies (Thomas et al. 2005, 2007). Since the results are similar, we
do not discuss results from NFW models.

Figure~5 plots the relationships between the four parameters for the
models. The parameter space is much larger than shown in this figure
since we only focus on the region around the best fit. In each panel
we include all models that are in that parameter range, irregardless
of their $\chi^2$ values. The color and symbol size, however,
indicates those models that are within the 68\% ($\Delta\chi^2=1$
after marginalizing over the other 3 parameters) and the 95\%
($\Delta\chi^2=4$) confidence bands of the best-fitted model. The idea
is to explore any correlations/degeneracies among the
parameters. There is not an obvious correlation between M/L and black
hole mass in the plotted region around the minimum $\chi^2$. However,
increasing the confidence band to 95\% shows a correlation. This
degeneracy is expected. Since the dynamical models only fit for the
enclosed mass, as the contribution from the stars is increased in the
central region, the contribution from the black hole has to
decrease. An obvious degeneracy within the 68\% band is that between
M/L and scale radius of the halo. As the scale radius is decreased,
for a given circular velocity, the contribution of the dark matter in
the central regions is increased, thereby decreasing the contribution
from the stars. These degeneracies are the essential reason why the
dark halo properties affect the inferred black hole mass in
M87. Beyond the plotting region, there are no points within the 68\%
confidence region, which allows us to set reasonable uncertainties on
the parameters.

\subsection{Gravitational Potential}

Figure~6 plots the gravitational potential. The black lines are the
potentials of the models that are within the 68\% limit
($\Delta\chi^2=1$ for the marginalized parameters). We also plot the
gravitational potential as inferred from the X-ray profile as
presented in Churazov et al. (2008). These have all been scaled to be
zero at 330\arcsec\ as in Churazov et al. The red line is the model
from Churazov et al. and it is clearly dissimilar from the best fit
potential derived from the stellar and cluster dynamics. The green
line is one of the models we used which is close to the gravitational
potential from the X-rays; we use that model to represent the X-ray
profile in subsequent analysis. Between the green line and the
best-fitted model, $\Delta\chi^2=18$, which implies the potential
derived from the X-ray is significantly poorer fit to the
kinematics. We find a potential that is deeper than the X-ray derived
potential. Churazov et al. (2008) explore a shock model for the X-rays
in an attempt to explain the wiggles seen in the X-ray potential, but
this model would not explain the large offset. Additional galaxies
with potentials derived from both stars and X-rays will be important
to study. This difference in the potential may be specific to M87.

\vskip 10pt \psfig{file=f6.ps,width=9cm,angle=0}
\figcaption[f6.ps]{The gravitational potential versus radius for M87.
The solid black lines are the dynamical models from this paper that
are within the 68\% confidence band of the best fit. The red noisy
line is the gravitational potential as derived from the X-ray gas
emission from Churazov et al. (2008). The green solid line is the
parameterization from our models that well matches the X-ray profile
(the parameters of the matched model are $6\times10^9\Msun$, 8.0,
800~\kms, 35~kpc for black hole mass, M/L$_V$, circular velocity and
core radius, respectively. The difference in $\chi^2$ between our
representation of the X-ray potential and that of the best fit is 18.
The units of the potential are as given in Churazov et al. and all
models are scaled to R=330\arcsec.
\label{fig6}}
\vskip 10pt

\subsection{Enclosed Mass and M/L Ratio}

Figure~7 plots the enclosed mass. The black lines represent the models
that are within the 68\% confidence limit. The enclosed mass flattens
in the center due to the black hole and rises linearly due to the dark
halo. The red line is the stellar contribution assuming M/L=5.0. The
green line is the enclosed mass from our model which best matches the
gravitational potential derived from the X-rays. In the potential
derived from the X-rays (green line in Figure~6), there is a
transition around 300\arcsec\ from the steep profile beyond that to a
shallow profile inside of that radius.  This transition manifests
itself as a wiggle in the enclosed mass profile around
200--300\arcsec. The enclosed mass derived from the stellar kinematics
instead show a very smooth gradient from a few arcseconds out to
1000\arcsec. If enclosed mass profiles are relatively smooth, as the
one derived here, it will be fairly easy to constrain that profile
using a non-parametric approach. This approach would be much cleaner
to implement since the kinematics only measure enclosed mass. By
trying to constrain the stellar contribution to the mass and the dark
halo contribution leads to degeneracies as seen in the above
Figures. However, this alternative approach would require some
simplification of the enclosed mass profile in order to use a
reasonable amount of cpu time (e.g., the four parameters explored
above required about 20,000 hours).

\vskip 10pt \psfig{file=f7.ps,width=9cm,angle=0}
\figcaption[f7.ps]{The mass profile for M87. The black lines represent
the models that are within the 68\% confidence band of the best fit
(as in Figure~6). The green line is the mass profile derived from our
representation of the X-ray gravitational potential (i.e., the green
line in Figure~6). The red line is the average contribution from the
stars, where we use the light profile in Figure~1 times 6.3 (the
best-fitted M/L). The mass profiles for the dynamical model show a
smooth transition from 30 to 1000\arcsec, whereas the X-ray profile
shows a kink.
\label{fig7}}
\vskip 10pt

\vskip 10pt \psfig{file=f8.ps,width=9cm,angle=0}
\figcaption[f8.ps]{The integrated M/L versus radius. The red dotted
lines represent the models that are within the 68\% confidence band of
the best-fitted potentials (black lines in Figure~6). The solid black
lines are the same models excluding the contribution from the black
hole. The green solid and dotted lines are our best representation of
the X-ray potential without and with a black hole. The horizontal blue
line is the M/L derived from stellar population models. From the red
dotted lines, there is no radial region in M87 that has a constant
M/L, since one always has to consider the contribution from either
black hole or dark halo.
\label{fig8}}
\vskip 10pt

Figure~8 plots the integrated mass-to-light ratio. We include all
models within the 68\% confidence limit of the best-fitted model. M/L
profiles for these models are shown as the red dotted lines in the
figure. The increase at small radius is from the black hole and the
increase at large radius from the dark halo. There is no region in M87
where the M/L is constant. The solid black lines in Figure~8 are the
M/L profiles including just the stars and dark halo, and are intended
to show the range of stellar M/L values. Figure~8 also plots the M/L
profile from our representation of the X-ray potential, both without
(solid green) and with (dashed green) a black hole. The blue
horizontal line is the best-fitted value of the stellar M/L from
stellar population models (Cappellari et al. 2006). The stellar
population models use Kroupa IMF with lower mass cut-off of
0.01. The stellar population models may have a large range of M/L
values depending on the lower mass cut-off. The adopted value of 0.01
is quite low and may bias the M/L's high if the lower cut-off is more
massive.

\section{Discussion}

There are a variety of results from this work, some determined more
robustly than others. Very robust results are 1) the black hole mass
in M87 from stellar dynamical measures is over 2 times higher when
including a dark halo, 2) the dynamical stellar M/L is 2x lower when
including a dark halo, and 3) uncertainties on parameters
significantly increase when including a dark halo. A model dependent
result from this work, although still very significant, is that 4) the
gravitational potential derived from X-rays for M87 is inconsistent
with that derived from kinematics.  Below we discuss each of these
points and their implications.

\subsection{Black Hole Mass}

One of the more robust results from this work is the increase in the
black hole due to inclusion of the dark matter halo. The reason is
quite simple: dynamical models only measure enclosed mass and if, in
the central regions, one lowers the stellar contribution with a
smaller M/L, then that mass must be compensated by a corresponding
increase in the black hole mass. For M87, since the light profile is
so shallow, the dynamically-derived M/L value will contain information
from stars that spend most of their time at large radii where the dark
halo dominates. Thus, excluding a halo will result in pushing the M/L
higher to match the observed velocity dispersions. One could partially
compensate for this by excluding kinematics at large radii in a
dynamical analysis. This may work somewhat for some galaxies, but as
seen in Figure~8, there is no region in M87 where there is a constant
M/L, causing a need to model simultaneously the black hole and the
dark halo.

A concern from this work could be that the dark halo parameters are
poorly constrained by the globular cluster kinematics, leading to a
biased result due to the above degeneracy. We have therefore run a
grid of models assuming the potential derived from the X-rays and
fitting for black hole mass and stellar M/L. We find nearly identical
results when fitting for the dark halo as well. The dark halo that we
derived is only moderately more massive than the one from X-rays, but
using either gives the same answer for the black hole mass. This is
reassuring as it leads to a clean result for the black hole
mass. Thus, we are confident that in M87 the black hole mass is
$6.4(\pm0.5)\times 10^9~\Msun$ derived from stellar kinematics.

There are significant implications for having a black hole mass this
large. There has long been an issue as to why black hole masses
derived from quasars can exceed $10^{10}~\Msun$, whereas the masses
derived from local galaxies never approach that. M87 and NGC4649
(Gebhardt et al. 2003, Humphrey et al. 2008) have the largest measured
black holes around $3\times10^9~\Msun$. The space density of quasars
with massive black holes imply that we should have many black holes
approaching $10^{10}$ out to about 100 Mpc (Shields et al. 2003). Many
studies have pointed out this problem (e.g., Richstone et al. 1998,
Sheilds et al. 2003, Lauer et al. 2007, Bernardi et al. 2007,
Salviander et al. 2008).  With a black hole mass of
$6.4\times10^9~\Msun$, the agreement between the space density of the
most massive black holes is in closer agreement with that derived from
quasars. Because this change is due to an issue in the modeling (and
not sample selection) and because only one object can have a such a
large effect, it is not worthwhile at this point to do a detailed
comparison. A much better approach would be to re-examine the other
massive galaxies and study additional ones.

Another implication of biased and larger black hole masses is the
effect it will have on correlations of the black hole mass with global
galaxy properties. The correlation of black hole with velocity
dispersion (Gebhardt et al. 2000, Ferrarese \& Merritt 2000) has led
to a large amount of theoretical research as to the cause of the
correlation and the value of the slope (e.g., Hopkins et al. 2007). If
the measured black holes at the upper are biased low, this will
obviously have a significant lever arm on the value of the slope and
scatter. The black hole masses were already high and possibly hint at
a curvature in the relation (Wyithe 2006, and see the recent analysis
of Gultekin et al. 2009). Correlations with other parameters would
also need re-evaluating. Again, it is not prudent to re-analyse the
current sample in regards to the correlations until the bias is
understood well with both re-examinations and larger samples.

There is a difference of the black hole mass measured with our
analysis compared to the previous HST gas kinematic measure.
Macchetto et al. (1997) use gas kinematics measured with HST to derive
a black hole mass of $3.2(\pm0.9)\times10^9~\Msun$. They use a
distance of 15 Mpc. Transforming to our assumed distance of 17.9 Mpc
gives a mass of $3.8(\pm0.9)\times10^9$. The radii of the gas emission
is in a region that is dominated by the black hole, so there is little
difference if the stellar M/L changes, as we measure in this
paper. For the black hole mass, $6.4\pm0.5$ compared to $3.8\pm0.9$ is
only a 2-sigma difference and while that may not be statistically
significant, it is important to consider possible causes for the
difference. There have been very few studies of galaxies with black
hole masses measured from both stars and gas. Of those that do have
both measures, the results can vary greatly: IC1459 has the gas
measure 8 times larger than that of the stars (Cappellari et al. 2002)
and CenA has the stellar measure 7 times larger than that of the gas
(Silge et al. 2005, Neumayer et al. 2007,), although Cappellari et
al. (2009) show that they are consistent for CenA.  The maser galaxy,
NGC4258, has a consistent mass estimate (Siopis et al. 2009), although
using the masing gas is not the same as using the gas further out. Ho
et al. (2002) discuss ways to optimize use of gas kinematics for
measuring the black hole mass by focussing on those galaxies with
well-organized dust lanes. For M87, the gas disk is not very well
defined (see Macchetto et al 1997) and the uncertainties with the
inclination can be large, but it would be difficult to have
inclination explain all of the difference. More significant concerns
with using the gas kinematics is that the intrinsic density
distribution is poorly constrained and the gas emission lines usually
show large velocity broadenings that are not generally considered. For
M87, Macchetto et al. include an estimate of the density distribution
but there are clearly large uncertainties since many volume density
distributions can project to give similar projected
densities. Macchetto et al. model the line centroids of the gas
emission to determine the gravitational potential. For their assumed
intrinsic gas distribution (which has a hole in the center), they also
match the large velocity dispersions seen in the emission lines;
however, the unknown intrinsic distribution can strongly effect the
gas centroids and dispersions.  The problem is that it is not clear
whether one should include the velocity broadening (thereby assuming a
hot component to the gas orbits) or exclude them (assuming some type
of turbulent motion), as discussed in Verdoes Kleijn et al. (2006).
For M87, this effect could account for some of the difference.

The resolution of the difference between the gas kinematics and the
stellar kinematics will best come from stellar kinematic data taken
with better spatial resolution. Whatever the value of the black hole
mass however, the dynamical models show a strong effect between
including and excluding a dark halo when the influence of the black
hole is not well resolved.

\subsection{Stellar M/L}

The stellar M/L$_V$ that we derive without a dark halo is 10 (9.7
correcting for extinction) and with a dark halo is 6.3 (5.3 correcting
for extinction). This is a very significant difference, but it is
obvious as to why that happens since the dynamical models simply are
trading enclosed mass in stars with enclosed mass in dark
matter. Adopting the gravitational potential as derived from the
X-rays requires a stellar M/L of 8, but this model provides a
significantly worse fit than our best fit. We compare to M/L as
derived from stellar population models. Cappellari et al (2006) use
Kroupa IMF to derive M/L$_I=3.33$ and find a dynamically-derived
M/L$_I=6.1$, using $A_B=0.1$ and distance of 15.6 Mpc. Transforming to
V-band and 17.9 distance implies M/L$_V=5.2$ (stellar population) and
M/L$_V=9.6$ (dynamical). Our dynamically-derived M/L is in good
agreement with that derived from the stellar population models. Also,
when we exclude the dark halo, we find a very similar M/L as the
dynamically-derived M/L from Cappellari et al., who do not include a
dark halo. Thus, the M/L derived from stellar population models
supports our conclusion that the dark halo is important for deriving
the proper M/L and, subsequently, the black hole mass.

The physical understanding of the Fundamental Plane is affected by the
dark matter fraction as a function of galaxy mass, which implies being
able to measure the stellar M/L. Until more galaxies are examined at
this level of detail (as in Thomas et al. 2007b), we do not attempt to
address implications for the Fundamental Plane.

\subsection{Parameter Uncertainties}

As we increase in the number of parameters in the dynamical models,
the uncertainties on those parameters will also increase. In fact,
there are some parameters which we have not explored for M87. Two
important ones include inclination and triaxiality. For inclination,
if M87 is spherical, then it should not have a large effect. However,
if M87 is very flattened and seen face-on, inclination will be
important; although since galaxies with the mass of M87 are all seen
as nearly round, the likelihood is that they are spherical. In
addition, M87 is flattened at large radii. Triaxiality is a bigger
concern, since these large galaxies tend to be triaxial in numerical
simulations (although no conclusive evidence yet either way from
observations). Triaxial models (van den Bosch et al. 2008) have only
recently been developed, and it will be important to study M87 as a
triaxial system. The uncertainties play a key role in understanding
the physical nature of the black hole correlations. Including a dark
halo already increased the uncertainties on the black hole and
particularly on the stellar M/L. Thus, until additional analysis on
the degeneracies is explored and a larger sample is included, it would
be prudent to include a modest systematic bias uncertainty for the
black hole mass estimates when studying the correlations and their
intrinsic scatter.

\subsection{Dark Halo and X-ray Properties}

It is generally assumed that the X-ray emitting gas can be used as a
robust tracer of the gravitational potential, and there have been
studies that show rough agreement between the two (as shown in
Churazov et al. 2008). However, all of the dynamical studies rely on
fairly restrictive assumptions about the stellar distribution
function: Wu \& Tremaine (2006) assume a two-integral distribution
function, McLaughlin (1999) assume spherical isotropic distribution
function, and both Kronawitter et al. (2000) and Romanowsky \&
Kochanek (2001) assume a spherical distribution. Given that the
stellar density profile at large radii is very flattened and that the
globular clusters show significant rotation, it is important to
explore more general models as the ones presented here. We find rough
agreement with the enclosed mass at some locations in M87, but the
shape of the potential profile is quite different. The gravitational
potential in Figure~6, in fact, shows fairly significant differences
between the two derived potentials. This figure, however, is not as
fair a comparison since the dynamical models directly measure the
enclosed mass and it is more instructive to focus on that aspect.
Since the enclosed mass is a derivative of the potential, one needs to
focus on the change in shape of Figure~6; for $R<40$\arcsec, the
derivative of the potential is very similar, which subsequently leads
to good agreement with the enclosed mass. At $R>600$\arcsec, there is
also good agreement in the enclosed mass (and derivative of the
potential). But it is the radii from $40<R<600$ where there is the
strongest disagreement, since the potential derived from X-rays shows
a strong kink there while that derived from stellar kinematics shows a
smooth transition. This region is where there is no stellar data with
modest coverage by globular cluster kinematics; however, given that we
are using parametric models for the dark matter profile, it is a
robust result that the X-ray potential is a significantly worse fit to
the data than the best-fitted logarithmic potential. It is clear
though that additional data at these radii are very much needed. We
note that the kinematics of Sembach \& Tonry (1996) are more
consistent with our best-fitted profile than with the X-ray profile;
however, the uncertainties are large and better data is needed to make
the comparison conclusive.

The dark halo parameters are, for a logarithmic potential, a circular
velocity of $715\pm15$~\kms and a core radius of
$14\pm2$~kpc. Kronawitter et al. (2000) finr a similar halo for
M87. This halo is also similar to those measured in Thomas et
al. (2007b) and Gerhard et al. (2001) in terms of the circular
velocity but M87 has a smaller core radius for galaxies of somewhat
smaller mass. We also ran a smaller grid of potentials that follow an
NFW profile. The best fitted parameters were around a concentration
of 15 and a scale radius of 50 kpc. We do not provide uncertainties
since we did not explore a very dense grid. Using either NFW or
logarithmic, the potential derived from X-rays was statistically
excluded.

M87 may not be the best case where X-ray gas emission provides an
adequate representation of the potential. The X-ray potential shows
significant deviations from a smooth curve, and Churazov et al. (2008)
use these deviations to derive a jet model, with cooling and heating
regions. This model, however, does not explain the difference at
$40<R<600$\arcsec. It would be important to study other galaxies that
do not show the strong deviations seen in M87, but it is possible that
the X-ray gas in M87 may not be an adequate tracer of the potential.

\subsection{Next Steps}

There are a variety of ways to move forward. On the data side, there
is a strong need for observations at small radii using observations
assisted with adaptive optics on a large telescope. The faint central
surface brightness of M87 made it not possible to observe with the
spectrographs on HST, but it is an ideal target for ground-based
telescopes. It should be straight-forward to detect a black hole with
as high a mass as presented in this paper. At larger radii, there is a
critical need to obtain stellar kinematics in the region from
$40<R<300$\arcsec. This is feasible using current ground-based
instruments (e.g., VIRUS-P on the 2.7~m telescope at McDonald
Observatory, Hill et al. 2008). The large radial data will strongly
limit dark halo models, and allow us to not rely as heavily on the
kinematics from the small number of globular clusters. Both of these
observational avenues, at small and large radii, are in progress.

On the modeling side, it would be worthwhile to include more general
models, such as triaxiality, and an exploration of inclination. For
this paper, we use a parametric form for the dark halo and there is
little theoretical justification for such an approach. A cleaner
approach would be to use a non-parametric estimate for the enclosed
mass, such as a mass profile defined in radial bins. This method would
directly include any change of mass-to-light ratio of the stars. The
enclosed mass is directly measured from the kinematics, and then
interpretation of how to split between stars and dark halo could come
subsequently. Given that the enclosed mass profile is relatively
smooth (Figure~7), it should be an easy matter to constrain it.
However, with just 4 parameters, it was already a large computer
effort, which could obviously increase dramatically as the number of
fitted mass bins increase.

There is a strong need to explore additional galaxies. It could be
that since M87 has no region where there is a constant M/L (see
Figure~8), that the degeneracies between black hole and dark halo,
mediated by the stellar M/L become very large. Galaxies with smaller
black holes and/or smaller relative fraction of dark halo should be
studied with this type of analysis.

Whether or not the dark halo parameters presented here correctly
reflect reality (we obviously agree that they do), we performed a
simple test of measuring the black hole with the same modeling code
including and excluding a dark halo. In a differential sense, the
importance of including the dark halo is one of the strongest and
cleanest results. All black holes measured, to date, including those
of the authors (Gebhardt et al. 2003) exclude a dark halo (but see the
discussion in Nowak et al. 2008). While we expect that M87 shows the
largest effect due to its shallow light profile, we are at the point
now where the details of the black hole correlations are important and
understanding all biases are therefore essential.

\acknowledgements

KG thanks the Max-Planck-Institut fuer Extraterrestrische Physik and
Carnegie Institute of Washington for their excellent support and
hospitality.  KG is very grateful to John Kormendy who has been
pushing for a few years to understand any possible degeneracy between
black hole mass and dark halo. We are grateful for conversations with
Ralf Bender, Douglas Richstone, Scott Tremaine, and Tod Lauer. We
would not have attempted this analysis without the
publically-available 2D dataset from SAURON; it is a tremendous
service to the community that the SAURON team choose to make the data
public and in an easy to use format. The project would not have been
possible without the excellent facilities at the Texas Advanced
Computing Center at The University of Texas at Austin, which has
allowed access to over 5000 node computers where we ran all of the
models. KG acknowledges NSF-CAREER grant AST03-49095.

\end{document}